\documentclass[aps,prl,showpacs,superscriptaddress,longbibliography,twocolumn,floatfix]{revtex4-1}
\usepackage{graphicx}
\usepackage{amsmath}
\usepackage{amssymb}
\usepackage{mathtools}
\usepackage{makeidx}
\usepackage{amsfonts}
\usepackage{bm}
\usepackage{dcolumn}
\usepackage{color}
\usepackage{xcolor}
\usepackage{xfrac}
\usepackage{wasysym}
\usepackage[colorlinks,
            linkcolor=blue,
            anchorcolor=blue,
            citecolor=blue,
urlcolor=blue
            ]{hyperref}

\newcommand{\be}{\begin{equation}}
\newcommand{\ee}{\end{equation}}

\begin{document}

\title{Speed limit for a highly irreversible process and tight finite-time Landauer's bound}

\author{Jae~Sung~Lee} \email{jslee@kias.re.kr}
\affiliation{School of Physics, Korea Institute for Advanced Study, Seoul 02455, Korea}
\author{Sangyun Lee}
\email{JSL and SL equally contributed to this work.}
\affiliation{School of Physics, Korea Institute for Advanced Study, Seoul 02455, Korea}
\author{Hyukjoon Kwon}
\email{hjkwon@kias.re.kr}
\affiliation{School of Computational Sciences, Korea Institute for Advanced Study, Seoul 02455, Korea}
\author{Hyunggyu~Park} \email{hgpark@kias.re.kr}
\affiliation{School of Physics, Korea Institute for Advanced Study, Seoul 02455, Korea}
\affiliation{Quantum Universe Center, Korea Institute for Advanced Study, Seoul 02455, Korea}

\date{\today}

\begin{abstract}
Landauer's bound is the minimum thermodynamic cost for erasing one bit of information. As this bound is achievable only for quasistatic processes, finite-time operation incurs additional energetic costs. We find a “tight” finite-time Landauer's bound by establishing a general form of the classical speed limit. This tight bound well captures the divergent behavior associated with the additional cost of a highly irreversible process, which scales differently from a nearly irreversible process. We also find an optimal dynamics which saturates the equality of the bound. We demonstrate the validity of this bound via discrete one-bit and coarse-grained bit  systems. Our work implies that more heat dissipation than expected occurs during high-speed irreversible computation.
\end{abstract}

 \maketitle

\emph{Introduction} --
Memory erasure is an elementary operation in irreversible computation. As the erasing operation incurs a thermodynamic cost and takes a finite physical time, low energy consumption and a short process time are critical requirements for efficient computation. The fundamental limitation of the energetic cost is given by Landauer's principle~\cite{Landauer, Parrondo_review}, which states that at least $k_{\rm B} T \ln 2$ of work is necessary to erase a single bit memory, where $k_{\rm B}$ is the Boltzmann constant and $T$ is the environment temperature. Landauer's bound is universal in the sense that it is independent of memory device type or physical platform. This bound has been confirmed experimentally using various physical setups, including a double-well potential realized by optical tweezers~\cite{Berut2012, Berut2015} and virtual potential~\cite{Jun2014, Gavrilov_2016}, an electric-circuit system~\cite{Orlov2012}, and a nanomagnetic memory bit~\cite{Hong2016, Martini2016}.

In real-world situations, however, this fundamental bound is less practical as it requires a quasistatic process, which far exceeds the system's relaxation time scale. The reported experimental time scale of the “quasistatic” erasure process ranges from a few hundred milliseconds~\cite{Dago} to several tens~\cite{Berut2015} or hundreds ~\cite{Jun2014} of seconds, which is far from the time required for practical computation. Therefore, it is important to understand the finite-time effect on thermodynamic cost, which generally increases as a process becomes faster and more irreversible \cite{Esposito2013PRE, Aurell2012, boyd2018, Paolo, Seifert2007PRL}. Several experimental studies have suggested that the minimum energetic cost should increase by an additional cost inversely proportional to the erasing time $\tau$, i.e., $k_{\rm B} T \ln 2 + C/\tau$, with a system-dependent constant $C$~\cite{Berut2012,Berut2015,Gavrilov_2016,Jun2014}. This behavior has also been investigated theoretically for the classical stochastic system described by the overdamped Langevin equation~\cite{Proesmans1, Proesmans2}, and an open quantum system described by the Lindblad equation~\cite{Vu2022PRL}.

{These studies suggest that  a trade-off relation plays a central role in understanding the overhead cost of the Landauer's bound. Over the last decade, various }types of trade-off relations have been reported in stochastic systems and also for open quantum systems
such as thermodynamic uncertainty relations~\cite{Shiraish2016, JSLee2020, Seifert2015PRL, DechantPNAS, Horowitz2020Review, Hasegawa2019PRL, JSLee2021, HasegawaQTUR2020, HasegawaQTUR2021},
kinetic uncertainty relations~\cite{Garrahan2017simple, Terlizzi2018kinetic, Hiura2021kinetic, Pal2021thermodynamic1, Pal2021thermodynamic2} and speed limits for state change~\cite{Shiraishi2018, Hasegawa2020PRE, Vu2021PRL, Ito2021PRL, PhysRevX.10.021056, Esposito2020, Nicholson2020, Wolpert2021, Dechant2022,delvenne2021tight}. Recently,
Zhen et al.~\cite{Zhen} showed that the $1/\tau$ behavior of the “minimum work bound” of erasing processes is
governed by the speed-limit inequality associated with the thermodynamic cost.

In this Letter, we first present a simpler way to derive the general form of the speed limit introduced in Ref.~\cite{delvenne2021tight}, which can have various functional forms. Two different speed limit regimes are considered in terms of the degree of irreversibility.
For a nearly reversible process, we retrieve the previous speed limits~\cite{Shiraishi2018, Hasegawa2020PRE} by taking a simple functional form, that provide a tight bound on the operation time in terms of entropy production (EP) and dynamical activity. However, this bound gradually loosens as the process becomes more irreversible. We find a tight bound for a highly reversible process from the general speed limit with a {different} functional form. In the limit of high irreversibility, this new bound becomes finite, depending solely
on dynamical activity.

We apply this general speed limit to calculate the tight bound of the additional cost associated with a finite-time erasing operation. We demonstrate that this bound scales as $1/\tau$ for a nearly reversible process, while much stronger divergence appears for a fast or highly irreversible process.
As a result, high-speed irreversible computation requires much more heat dissipation and thus associated supporting cooling architecture.  We also find an explicit protocol which meets the equality condition of the bound. 


\emph{Speed limit for a highly irreversible process} --
Suppose that the time evolution of a probability distribution $\boldsymbol{p} (t) = \{ p_n(t) \}$ at time $t$ is described by
$$
\dot p_n(t) = \sum_{m} R_{nm}(t) p_m(t),
$$
where the transition rate matrix $\boldsymbol{R}(t)$ satisfies the condition $R_{nm}(t) \geq 0$ for $m\neq n$ and $R_{nn}(t) = -\sum_{m (\neq n)} R_{mn}(t) $. The statistical distance between the initial and the final distributions after time $\tau$ can be measured by the total variational distance
$$
\ell = d_T(\boldsymbol{p}(\tau), \boldsymbol{p}(0)) \equiv  \frac{1}{2} \sum_n \left | p_n(\tau) - p_n(0) \right|.
$$
We first establish a general form of the speed limit, in terms of the distance $\ell$, the EP $\Sigma^*$, and the total activity (number of jumps) $A_{\rm tot}$ during $\tau$, given by
\begin{equation}
 \frac{\ell}{A_{\rm tot}}  \leq f\left( \frac{\Sigma^*}{ A_{\rm tot}}  \right),
\label{eq:speed_limit1}
\end{equation}
with an appropriate choice of a monotonically increasing  concave function $f$ as listed in Table~\ref{tab:f_choices}. Hereafter, we set
$k_{\rm B}=1$ for convenience. The EP $\Sigma^* = \int_0^\tau (\dot{\Sigma}^*)dt$ is characterized by the transition matrix of the adjoint process $\boldsymbol{R}^*(t)$~\cite{Esposito2010, Kurchan2012, Spinney2012, HyunKeun2013}, where the corresponding EP rate is defined as
\begin{equation}
\dot{\Sigma}^* \equiv \sum_{n\neq m} R_{nm}(t) p_m(t) \ln\left[ \frac{R_{nm}(t)p_m(t)}{R^*_{mn}(t) p_n(t)}\right]. \label{eq:gen_EP}
\end{equation}
We assume that the adjoint process is also stochastic with the same {\em escape} rate; $R^*_{nn}(t) = -\sum_{m (\neq n)} R^*_{mn}(t) = R_{nn}(t)$. A trivial choice of the adjoint process is taking it the same as the time-reversal process, $R^*_{mn}(t) = R_{mn}(t)$, which leads to the total EP $\Sigma$, i.e. $\Sigma^* = \Sigma$. Alternatively, by considering the instantaneous steady-state $\boldsymbol{p}^{\rm ss}(t)$ such that $\dot{\boldsymbol{p}}^{\rm ss}(t) = 0$, the adjoint process defined as $R^*_{mn}(t) = R_{nm}(t) \left( \frac{p^{\rm ss}_m(t)}{p^{\rm ss}_n(t)} \right)$ leads to the Hatano-Sasa (excess) EP $\Sigma^* = \Sigma_{\rm HS}$. Since $0\leq \Sigma_{\rm HS} \leq \Sigma$~\cite{Sasa2001,Esposito2010}, $\Sigma_{\rm HS}$ always gives a tighter bound than $\Sigma$ in Eq.~\eqref{eq:speed_limit1}.

To derive Eq.~\eqref{eq:speed_limit1}, we first note that
\begin{equation}
\ell = \frac{1}{2} \sum_n \left | p_n(\tau) - p_n(0) \right| \leq \frac{1}{2} \int_0^\tau dt \sum_n |\dot{p}_n(t)|, \label{eq:derivation1}
\end{equation}
from the triangle inequality. Using $R^*_{nn}(t) = R_{nn}(t)$, the instantaneous change of the probability distribution is bounded by
\begin{align}
	\sum_n |\dot{p}_n(t)|
	&=\sum_n \left| \sum_{m(\neq n)} R_{nm}(t) {p}_m(t) + R_{nn}(t) {p}_n(t) \right| \nonumber  \\
	&=\sum_n \left| \sum_{m(\neq n)} \left\{ R_{nm}(t) {p}_m(t) - R^*_{mn}(t) {p}_n(t) \right\} \right|  \nonumber\\
	&\leq \sum_{n\neq m} \left| R_{nm}(t) {p}_m(t) - R^*_{mn}(t) {p}_n(t) \right|  \nonumber \\
	&= 2 A(t) d_T(\boldsymbol{Q}(t), \boldsymbol{Q}^*(t)), \label{eq:derivation2}
\end{align}
where $A(t) = \sum_{n\neq m} R_{nm}(t) p_m(t) = \sum_{m \neq n} R^*_{mn}(t) p_n(t)$ is  a jump rate at time $t$ and $\boldsymbol{Q}(t)$ ($\boldsymbol{Q}^*(t)$) denotes the normalized conditional joint probability distribution of the forward (reverse) process which is defined as follows:
$$
\begin{aligned}
	Q_{mn}(t) &=  P[m,n | {\rm jump}] = \frac{ (1-\delta_{nm}) R_{nm}(t) p_m(t)}{A(t)},  \\
	Q^*_{mn}(t) &= P^*[n,m | {\rm jump}] = \frac{ (1-\delta_{mn}) R^*_{mn}(t) p_n(t)}{A(t)}~.
\end{aligned}
$$
Thus, $d_T(\boldsymbol{Q}(t), \boldsymbol{Q}^*(t))$ captures how much irreversible the process is at time $t$. We remark that $A_{\rm tot} = \int_0^\tau dt A(t)$ has a meaning of the total number of jumps during the entire process. By combining Eqs.~\eqref{eq:derivation1} and \eqref{eq:derivation2}, we have the following inequality:
\begin{equation}
\ell \leq \int_0^\tau dt A(t) d_T(\boldsymbol{Q}(t), \boldsymbol{Q}^*(t)).
\label{eq:length_ineq}
\end{equation}

\begin{table}[t]
\caption{\label{tab:f_choices} Various choices of the concave function $f(x)$ satisfying Eq.~\eqref{eq:dist_bounds} and its inverse (convex) function $h(v) = f^{-1}(v)$.}
\begin{ruledtabular}
\begin{tabular}{ccc}
~ & $f(x)$ & $h(v)$  \\
Pinsker~\cite{Pinsker60} & $\sqrt{(x/2)}$ & $2v^2$  \\
Bretagnolle--Huber~\cite{Huber79} & $\sqrt{1-e^{-x}}$ & $-\ln(1-v^2)$ \\
Vajda~\cite{Vajda70} & n/a \footnote{Analytic compact expression is not available.} & $\ln\left(\frac{1+v}{1-v}\right)-\frac{2v}{1+v}$ \\
Gilardoni~\cite{Gilardoni08} & n/a $^{\rm a}$ & $\ln\left[\frac{(1+v)^{-1+v}}{1-v}\right]$ \\
Symmetric KLD \footnote{This bound is valid only when the KLD is symmetric.}~\cite{Gilardoni2008,gilardoni2006minimum}  & { n/a $^{\rm a}$ } & { $v \ln\left[\frac{1+v}{1-v}\right]$ } \\
\end{tabular}
\end{ruledtabular}
\end{table}
It is worth nothing that the EP rate can be expressed in terms of the conditional joint distributions as
\begin{equation}
\dot\Sigma^* =  A(t) D(\boldsymbol{Q}(t) || \boldsymbol{Q}^*(t)),
\label{eq:ent_rate_rel}
\end{equation}
where $D(\boldsymbol{p}||\boldsymbol{q}) \equiv \sum_x p_x \ln (p_x/q_x)$ is the Kullback-Leibler divergence (KLD) between two probability distributions $\boldsymbol{p}$ and $\boldsymbol{q}$.  Note that the KLD corresponding to the total EP rate ($\dot{\Sigma}$) is symmetric ($D(\bm p || \bm q) = D(\bm q || \bm p)$), while the KLD  is generally asymmetric for other choices such as $\dot{\Sigma}_{\rm HS}$.
There exist various choices of a monotonic concave function $f$ (see Table~\ref{tab:f_choices}) that connects the total variational distance and the KLD to obey the following inequality~\cite{Pinsker60, Huber79, Vajda70, Gilardoni08}:
\begin{equation}
d_T(\boldsymbol{p}, \boldsymbol{q}) \leq f(D(\boldsymbol{p} || \boldsymbol{q})).
\label{eq:dist_bounds}
\end{equation}
The speed limit is obtained by plugging in Eqs.~\eqref{eq:ent_rate_rel} and \eqref{eq:dist_bounds} to Eq.~\eqref{eq:length_ineq}, and then dividing both sides with $A_{\rm tot}$, which leads to
\begin{equation}
\frac{\ell}{A_{\rm tot}} \leq \frac{\int_0^\tau dt A(t) f\left( \frac{\dot\Sigma^*}{ A(t)}  \right)}{\int_0^\tau dt A(t)} \leq f\left( \frac{\Sigma^*}{ A_{\rm tot}}  \right), \label{eq:pre_speedlimit}
\end{equation}
from the concavity of $f$.

{ As $g(x) \equiv h(x)/(2x)$ is a monotonically increasing function for all $h$'s in Table~\ref{tab:f_choices}, where $h = f^{-1}$,
we can rewrite Eq.~\eqref{eq:pre_speedlimit} as
\begin{equation}
	\tau \geq  \frac{\ell}{ \langle A\rangle_\tau g^{-1} \left( \frac{\Sigma^*}{2 \ell} \right)} ,
	\label{eq:gen_speedlimit}
\end{equation}
by defining $\langle A \rangle_\tau \equiv A_{\rm tot}/\tau$ and $g^{-1}(x)$ the inverse function of $g(x)$.} Equation~\eqref{eq:gen_speedlimit} is the general form of the speed limit, where various types of bounds can be obtained based on the choice of { $h(x)$.}  The previous speed limit $\tau \geq 2  \ell^2 /(\langle A\rangle_\tau \Sigma^*)$ in Refs.~\cite{Shiraishi2018,Hasegawa2020PRE} is readily obtained by taking { $h(x)=x^2$ } (Pinsker~\cite{Pinsker60}), which is tight only for a nearly reversible (slow) process but yields a very loose bound for a highly irreversible process.
We note that for any $\Sigma^* \geq 2\ell$, this bound is even worse than the {\em fundamental} bound  $\tau \geq \ell/\langle A\rangle_\tau$ obtained from the minimum activity to change the probability distribution regardless of the EP, $A_{\rm tot} = \langle A \rangle_\tau \tau \geq \ell$~\cite{delvenne2021tight}.

We find that a speed limit can be tightened for a highly irreversible process with alternative choices of { $h(x)$} such as Bretagnolle--Huber, Vajda, Gilardoni, { and symmetric KLD } as listed in Table~\ref{tab:f_choices}.
All these four functions provide speed limits, always tighter than the fundamental bound,  which can be accessible only when $\Sigma^* \rightarrow \infty$. Therefore, in the highly irreversible limit, time is bounded solely by the dynamical activity, but not the EP. { The symmetric KLD bound is always the tightest among all $h(x)$, though it is valid only for the symmetric KLD.  Otherwise, the Gilardoni bound is the tightest
 for $\Sigma^* /\ell \geq 1.14$, while the Pinsker bound is the tightest elsewhere. Simple derivation of the symmetric KLD bound is presented in Supplemental Material (SM)~\cite{SM}. }

%
\emph{Tight finite-time Landauer's bound} --
The  speed limit in Eq.~\eqref{eq:speed_limit1} can be rearranged to bound the EP as

\begin{equation}
	\Sigma^* \geq \ell   \frac{h(v)}{v}
	\equiv B_H, \label{eq:gen_Landauer1}
\end{equation}
where  $v = \ell/A_{\rm tot}$ is the average distance change per jump, which ranges from $0$ to $1$, measuring the irreversibility of the process. When $v$ is close to $0$ ($1$), the distribution changes gradually (abruptly), so the process is nearly reversible (highly irreversible). The bound $B_H$ monotonically increases with $v$ for all $h$'s, where $H$ denotes a specific functional form, e.g., $H={\rm P}$ (Pinsker) and $H={\rm S}$ (symmetric KLD) { with $B_{\rm P}=2 \ell v$ and $B_{\rm S}=2 \ell\tanh^{-1} v$.}

Now, we use the EP bound to estimate the minimum cost for a finite-time erasing process. Suppose an erasing operation resets a one-bit system composed of $0$ and $1$ states with the associated probabilities $p_0(t)$ and $p_1(t)$, respectively. Let us assume that the initial bit is random with probability distribution as $(p_0(0), p_1(0)) = (1/2, 1/2)$, and the erasing process yields the final distribution $(p_0(\tau), p_1(\tau)) = (1-\epsilon, \epsilon)$ with  erasing error $\epsilon$ after time $\tau$. The statistical distance between the initial and final states becomes $\ell = 1/2- \epsilon$, and the Shannon entropy change of the system can be computed as $\Delta S_{\rm sys} = -\ln 2 -(1-\epsilon) \ln (1-\epsilon) - \epsilon \ln \epsilon$.
Furthermore, by setting $\Sigma^*$ as the total EP~\cite{note1}, we have $\Sigma^* = \Delta S_{\rm sys} + Q/T$, where $Q$ is the heat dissipated into the surrounding environment with temperature $T$ during the erasing process. {
As the total EP $\Sigma$ corresponds to the symmetric KLD,  we use the symmetric KLD bound
to obtain the tightest Landauer's bound.}

In the perfect erasing limit $\epsilon \rightarrow 0$, we get $\ell\rightarrow 1/2$ and  $\Delta S_{\rm sys} \rightarrow  -\ln 2$
and the finite-time Landauer's bound from Eq.~\eqref{eq:gen_Landauer1} is expressed as
\begin{equation}
	\frac{Q}{ T} \geq   \ln 2 + B_{\rm S} =\ln 2 +\tanh^{-1} v,  \label{eq:gen_Landauer2}
\end{equation}
{ where $B_{\rm S} $ represents the additional cost due to finite-time operation.
For small $v$ (nearly reversible), $B_\textrm{S}\simeq v= 1/(2\tau \langle A \rangle_\tau)$, which corresponds to the previously known $1/\tau$ behavior~\cite{Berut2012,Berut2015,Gavrilov_2016,Jun2014,Zhen}.  As we approach $v=1^-$ (highly irreversible regime),
$B_\textrm{S}$ diverges asymptotically as }
\begin{equation}
	B_\textrm{S} \simeq -  \frac{1}{2} \ln \left( 1- v \right) = - \frac{1}{2} \ln \left( 1- \frac{1}{2\tau \langle A \rangle_\tau} \right). \label{eq:finite-time_Landauer_cost}
\end{equation}
This implies that much higher dissipation should occur in a highly irreversible erasing operation.

Practical computation requires a small erasing error as well as a short operation time. To this end,
the transition rate from $1$ to $0$ state has to be large for fast operation, necessitating large driving (large $\langle A \rangle_\tau$).
In comparison, the reverse transition ($0$ to $1$) should be suppressed to prevent erasing-error operations. Therefore, the best strategy for a desired erasing operation is that all ``particles'' initially located at state $1$ jump to state $0$ once, and
no jump occurs afterwards; this condition can be read as $A_{\rm tot} \approx 1/2$ with $p_1(0)=1/2$.
Consequently, the operation for a practical erasing process should be highly irreversible with $v=\ell/A_{\rm tot} \approx 1$.
Thus, Eq.~\eqref{eq:finite-time_Landauer_cost} for a highly irreversible process is well deserved  for practical computation.

{ We find explicitly the {\em optimal} dynamics which minimizes the EP, satisfying the equality of Eq.~\eqref{eq:gen_Landauer1} with $h=h_{\rm S}$. Its sufficient condition is
\begin{equation}
	\frac{R_{01}(t) p_1(t)}{R_{10}(t) p_0(t)} = c ~({\rm const.}),~~~\forall ~0\leq t \leq \tau~,
	\label{eq:saturation_cond}
\end{equation}
along with monotonic change of $p_n(t)$ in time. The detailed derivation is presented in SM~\cite{SM}, where we also show that a process with $v \approx 1^-$ ($v \approx 0$) is realized with large $c$ and short $\tau$ ($c \approx 1$ and long $\tau$).  }

The finite-time Landauer's bound, Eqs.~\eqref{eq:gen_Landauer1} and \eqref{eq:gen_Landauer2}, is also applicable to a bit system made by coarsening, such as a Langevin system with a double-well potential~\cite{Berut2012, Berut2015, Gavrilov_2016, Jun2014}. This can be verified by the fact that the EP of a coarse-grained bit system is equal to or smaller than that of its original system without coarse-graining. See Eq.~(10) in Ref.~\cite{Zhen} and SM~\cite{SM} for details. Therefore, the original EP is also bounded by the same additional cost term in Eq.~\eqref{eq:gen_Landauer1}.

\emph{Numerical confirmation} --
\begin{figure}[!t]
\centering
\includegraphics[width=0.47\textwidth]{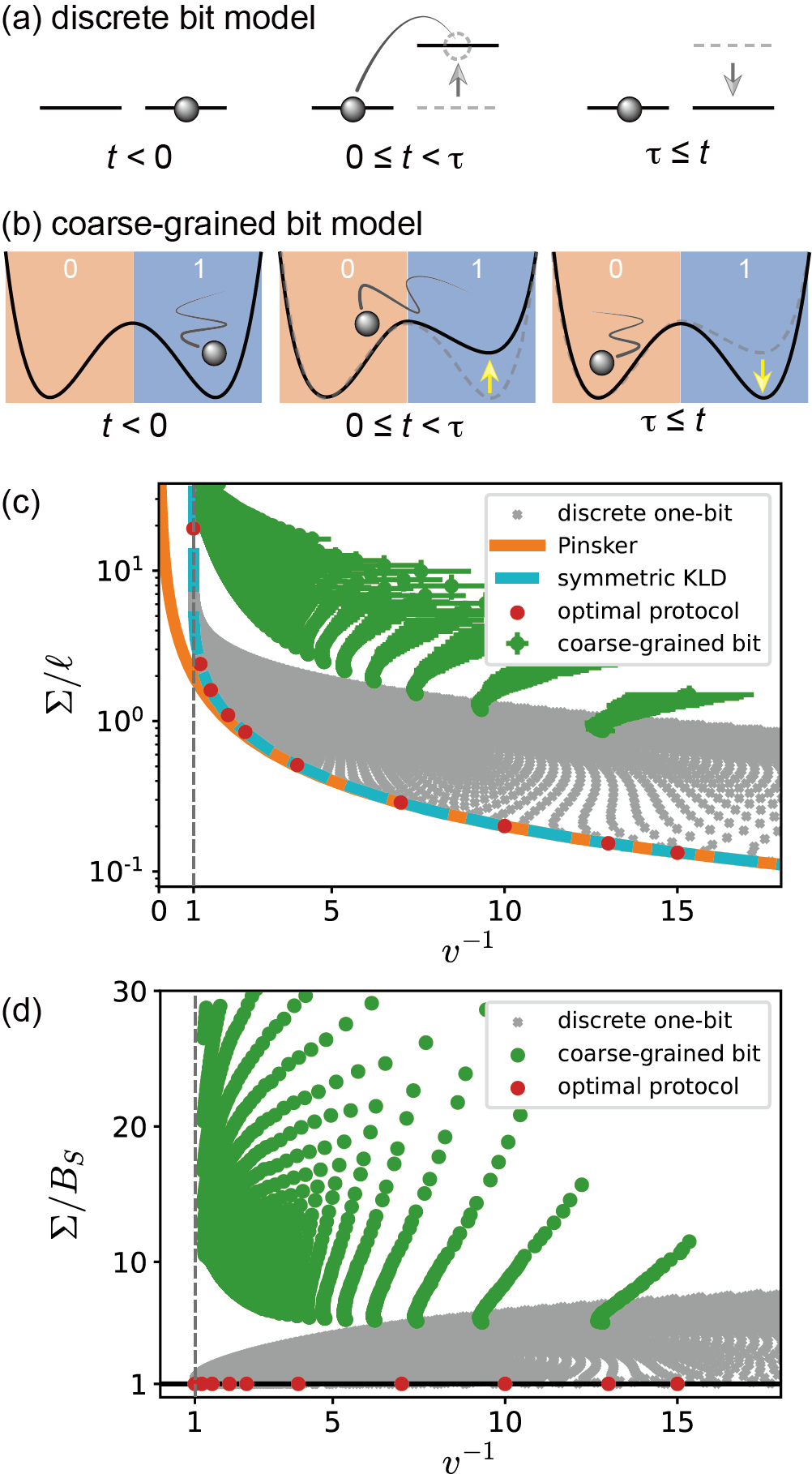}
\vskip -0.1in
\caption{ (a, b) Schematics of the two erasure models: discrete one-bit system (a) and coarse-grained bit system (b). (c) Plot of $\Sigma/\ell$ versus $v^{-1}$ for discrete system (gray $\times$) and continuous system  (green \newmoon).
The orange solid (sky-blue dashed) curve denotes the Pinsker (symmetric KLD) bound. { The result of the optimal protocol is denoted by red dot \newmoon.} (d) Plot of $\Sigma$ divided by the the symmetric KLD bound~$B_\textrm{S}$ versus  $v^{-1}$. { For a fixed $\ell$ and $A(t)$, $v^{-1}$ can be simply regarded as a scaled time.} The same data are used for (c) and (d). }
\label{fig:simul}
\vskip -0.1in
\end{figure}
Here we investigate two examples. The first one is a discrete one-bit system consisting of states $0$ and $1$ with energy levels $E_0 =0$ (fixed) and $E_1 (t)$ (time-varying), respectively. Its dynamics is described by the following master equation:
\begin{align}
	\dot p_0(t) =& R_{01}(t) p_1(t) - R_{10}(t)p_0(t),\nonumber \\
	\dot p_1(t) =& R_{10}(t) p_0(t) - R_{01}(t) p_1(t).
\end{align}
The transition rate $R_{nm}(t)$ satisfies the detailed balance condition, that is, with $\gamma(t) \equiv e^{- E_1(t)/T }/(1+e^{- E_1 (t) /T })$
\begin{align}
	R_{10}(t) = \mu(t) \gamma(t),~~	R_{01}(t) = \mu(t) (1-\gamma(t)),
\end{align}
where $\mu(t)$ is an overall transition rate. Erasing process is illustrated in Fig.~\ref{fig:simul}(a). The system is prepared with the initial distribution $(p_0, p_1) = (1/2,1/2)$ with $E_1 =0$ and $\mu =0$ for $t<0$. Here, $\mu =0$ indicates that the transition is blocked. $E_1$ and $\mu$ are abruptly raised to $E_{\rm eras} $ and $\mu_{\rm eras}$ at time $t=0$, respectively, and maintained up to $t=\tau$. And then, both $E_1 $ and $\mu $ are immediately lowered to $0$ at $t=\tau$. The final distribution at $t=\tau$ is $(p_0, p_1) = (1-\epsilon, \epsilon)$. This protocol is the simplest one in the $n$-step energy-raising procedure~\cite{Browne2014, Zhen}. The exact solution of this model is
\begin{align}
	p_1(t) = e^{-\mu t } p_1(0) +(1-e^{-\mu t}) p_{1,E_{\rm eras}}^{{\rm eq}},
	\label{eq:p1}
\end{align}
where $p_{1,E_{\rm eras}}^{\rm eq}= 1/[1+\exp (E_{\rm eras} /T) ]$. Using Eq.~\eqref{eq:p1}, we explicitly calculate the entropy change of the system $\Delta S_{\rm sys} = -\sum_i[ p_i(\tau) \ln{p_i(\tau)} - p_i(0)\ln{p_i(0)}] $ and heat $Q =  E_{\rm eras} [p_1(\tau ) -p_1(0) ]$, which leads to the total EP $\Sigma = \Delta S_{\rm sys} + Q/T$, as well as the total activity $A_{\rm tot}$.
{ We also construct an optimal time-dependent control of $R_{nm}(t)$ satisfying the saturation condition Eq.~\eqref{eq:saturation_cond}, of which the explicit form can be found in SM~\cite{SM}.}

The second example is a coarse-grained bit system consisting of a one-dimensional Brownian particle trapped in a double-well potential. Dynamics of the particle is governed by the following overdamped Langevin equation:
\begin{align}
	\gamma \dot x = -\frac{\partial V_{\rm DW}(x,t)}{\partial x} + \sqrt{2 \gamma  T}\xi(t),
\end{align}
where $x$ is position of the particle, $\xi(t)$ is a Gaussian white noise satisfying $\langle \xi(t) \xi(t')\rangle = \delta(t-t')$, and the double-well potential $V_{\rm DW} (x,t)$ is given as
\begin{align}
	V_{\rm DW}(x,t) =& E_b \left[ \left( \frac{x}{x_m} \right)^4 - 2 \left( \frac{x}{x_m} \right)^2  \right] + \Theta(t) \frac{x}{x_m},
\end{align}
where $\Theta(t)$ provides a time-dependent protocol. This model corresponds to the experimental setup in Ref.~\cite{paneru2021new}. The system can be treated as a coarse-grained bit memory by regarding the system being in state ``0'' (``1'') when $x \leq  0$ ($x > 0$). Then, the probabilities for the coarse-grained state $i$ ($i \in \{ 0, 1\}$) are
\begin{align}
	p_0^{\rm cg} (t) = \int_{x \leq 0 } dx P(x,t)  ~~\textrm{and}~~ p_1^{\rm cg} (t) = 1 - p_0^{\rm cg} (t),
	\label{eq:coarse_state}
\end{align}
where $P(x,t)$ is the probability distribution of the original continuous system. The erasing process of this model is presented in Fig.~\ref{fig:simul}(b). An initial state is prepared as the equilibrium state determined by the double-well potential with $\Theta(t)=0$ for $t<0$. As the potential is symmetric with respect to $x=0$,  $p_0^{\rm cg} (0) = p_1^{\rm cg} (0) =1/2 $.  At $t=0$, we immediately raises $\Theta(t)$ to $\Theta_{\rm eras}$ and maintain it up to $t=\tau$. $\Theta(t)$ then returns to $0$ at $t=\tau$. Due to the nonlinearity of the potential force, an analytic solution is not available. Instead, the total EP $\Sigma = \Delta S_{\rm sys}+Q/T$ is estimated by numerically  evaluating $\Delta S_{\rm sys}=- \int dx  [ P(x,\tau) \ln{P(x,\tau) }- P(x,0)\ln{P(x,0) } ] $ and $Q =\int_0^\tau dt (- \partial_x V_{\rm DM}) \circ \dot x (t)$, where $\circ$ is the Stratonovich product. $\ell $ and $A_{\rm tot}$ are estimated by using $p_i^{\rm cg}(t)$ and by counting the number of transitions between the different coarse-grained states~\cite{activity}.

Figure~\ref{fig:simul}(c) shows the plot of $\Sigma/\ell$ against $v^{-1}$ for the discrete and the coarse-grained bit models. The data for the discrete model are obtained by varying parameters $E_{\rm eras}$ and $\tau$ within the ranges $10^{-5} \leq E_{\rm eras} \leq 10$ and $10^{-10} \leq \tau \leq 20$ with fixed $\mu_{\rm eras}=1$ and $ T=1$. The data of the coarse-grained bit model are the simulation results for the parameter ranges (used in real experiment~\cite{paneru2021new}) of $0.1 k_{\rm B}T \leq \Theta_{\rm eras} \leq 10 k_{\rm B}T$ and $0.1~{\rm ms}\leq \tau \leq 110~{\rm ms}$ with fixed $x_m = 50~\text{nm}$, $k_{\rm B} T = 4.1~ \text{pN}\cdot\text{nm}$ ($T= 300~{\rm K}$), $E_b = 3k_{\rm B}T$, and $\gamma = 24\sqrt{2} k_{\rm B}T\cdot \text{ms}/\pi x_m^2$. Each point of the coarse-grained bit model in the plot is obtained by averaging $10^6$ realizations. The Pinsker and { the symmetric KLD bounds are presented along with the
result of the optimal erasing process in the figure and the comparison with other bounds is shown in SM~\cite{SM}.}

Indeed, the { symmetric KLD} tightly bounds the EP of the discrete bit model for all $v$. This tightness can be also checked in Fig.~\ref{fig:simul}(d), which presents the total EP divided by $B_\textrm{S}$ (see Eq.~\eqref{eq:gen_Landauer1}).
Note that the Pinsker bound is quite tight for nearly-reversible processes (small $v$); however, it becomes extremely loose near $v=1$. 
The data of the coarse-grained model are also well bounded by $B_{\rm S}$. However, the bound is not tight due to the ``intra EP'' induced by transitions  between {\em microstates} inside the same coarse-grained state. The detailed explanation is presented in SM~\cite{SM}. Thus, it is also important to reduce the intra EP for lowering the thermodynamic cost for a coarse-grained system.

\emph{Conclusion} -- We find the finite-time Landauer's bound, which is tight for { an erasing process with any irreversiblility and any error rate, from the general form of the speed limit. We also find an optimal dynamics which saturates the equality of the bound.}
This bound is applicable to a coarse-grained bit system as well as an intrinsically two-state system. We demonstrate that, for a highly irreversible process, the diverging behavior of the additional cost is much steeper than that of a nearly reversible process. This indicates that, in a practical computation, which belongs to a highly irreversible regime, reducing the operation time and error rate gives rise to much more heat dissipation than expected. Thus, enhancing the cooling power or heat tolerance of a memory device to maintain a proper device temperature is more critical when computation becomes more irreversible. Our formula is also directly applicable for estimating the proper bound of cooling power for a given computation speed, and the density of memory. Furthermore, to save thermodynamic costs, it is important to reduce the dissipation produced inside the same coarse-grained state. Subsequent experimental studies in various physical systems are anticipated in the future.


JSL and SL equally contributed to this work. Authors acknowledge the Korea Institute for Advanced Study for providing computing resources (KIAS Center for Advanced Computation Linux Cluster System). This research was supported by NRF Grants No.~2017R1D1A1B06035497 (H.P.), and individual KIAS Grants No.~PG064901 (J.S.L.), PG081801 (S.L.),CG085301 (H.K.), and QP013601 (H.P.) at the Korea Institute for Advanced Study.






\bibliography{speed}

\begin{thebibliography}{59}
\bibitem [{not()}]{note1}%
  \BibitemOpen
  \href@noop {} {\ }\bibinfo {note} {Note that for a two-state model, where
  $R_{nm}(t) p^{\rm ss}_m(t) = R_{mn}(t) p^{\rm ss}_n(t)$ is satisfied, the
  Hatano-Sasa EP coincides with the total EP, i.e. $\Sigma_{\rm HS} = \Sigma$;
  it can be easily checked by plugging in $R^*_{mn}(t) = R_{nm}(t) \left(
  \frac{p^{\rm ss}_m(t)}{p^{\rm ss}_n(t)} \right) = R_{mn}(t) $ into
  Eq.~\eqref{eq:gen_EP}.}\BibitemShut {Stop}%
\bibitem [{act()}]{activity}%
  \BibitemOpen
  \href@noop {} {\ }\bibinfo {note} {Note that the calculation of $A_{\rm tot}$
  from the original (not-coarse-grained) dynamics leads to the meaningless
  result as $A(t)$ diverges in the Langevin dynamics with a continuous
  space.}\BibitemShut {Stop}%
\end{thebibliography}

\end{document}


\title{Supplemental material for ``Speed limit for a highly irreversible process and tight finite-time Landauer’s bound''}

\author{Jae~Sung~Lee}
\affiliation{School of Physics, Korea Institute for Advanced Study, Seoul 02455, Korea}
\author{Sangyun Lee}
\affiliation{School of Physics, Korea Institute for Advanced Study, Seoul 02455, Korea}
\author{Hyukjoon Kwon}
\affiliation{School of Computational Sciences, Korea Institute for Advanced Study, Seoul 02455, Korea}
\author{Hyunggyu~Park}
\affiliation{School of Physics, Korea Institute for Advanced Study, Seoul 02455, Korea}
\affiliation{Quantum Universe Center, Korea Institute for Advanced Study, Seoul 02455, Korea}

\newcommand{\red}[1]{{\color{red}#1}}

\maketitle

\section{Simple derivation of the symmetric KLD (Kullback-Leibler divergence) bound}

The total variational distance is bounded by the Le Cam distance $d_{LC} (\bm p, \bm q)$ as follows:
\begin{align}
	d_T (\bm p, \bm q)^2  &= \left(\frac{1}{2} \sum_n \frac{|p_n - q_n|}{\sqrt{p_n + q_n}} \cdot \sqrt{p_n + q_n} \right)^2  \leq \frac{1}{4} \sum_n \frac{|p_n - q_n|^2}{p_n + q_n} \cdot 2 \equiv d_{LC} (\bm p, \bm q)^2. \label{eqS:LC}
\end{align}
The Cauchy–Schwarz inequality is used to derive the inequality of Eq.~\eqref{eqS:LC}.  By using the equality $|p-q|/(p+q) = \tanh |\frac{1}{2}\ln \frac{p}{q}|$ for $p,q \geq 0$ and the normalized probability $\tilde{p}_n \equiv \frac{|p_n - q_n|}{2d_{T} (\bm p, \bm q)}$, we can show that
\begin{align}
	d_{LC} (\bm p, \bm q)^2  &= d_{T} (\bm p, \bm q) \sum_n \tilde{p}_n \cdot \frac{|p_n - q_n|}{p_n + q_n} \nonumber \\
	& \leq d_T (\bm p, \bm q) \tanh \left( \sum_n \tilde{p}_n  \left| \frac{1}{2} \ln \frac{p_n}{q_n} \right| \right) \nonumber \\
	&=  d_T (\bm p, \bm q) \tanh \left[ \frac{1}{4 d_T(\bm p, \bm q)}  \sum_n (p_n - q_n) \ln \frac{p_n}{q_n} \right]  \label{eqS:LC1}
\end{align}
The concavity of the $\tanh$ function is used for the inequality of Eq.~\eqref{eqS:LC1}.
Combining Eqs.~\eqref{eqS:LC} and \eqref{eqS:LC1} and using the symmetrized KLD, $ \frac{1}{2} \sum_n (p_n - q_n) \ln \frac{p_n}{q_n} =  \frac{1}{2} \left[ D(\bm p || \bm q) + D(\bm q || \bm p) \right] \equiv D_{\rm S} (\bm p || \bm q) $, we have
\begin{align}
	2 d_T (\bm p, \bm q) \tanh^{-1} d_T (\bm p, \bm q) \leq D_{\rm S} (\bm p,\bm q).  \label{eqS:sKLDbound}
\end{align}
When the KLD is symmetric, that is, $D(\bm p, \bm q) = D(\bm q, \bm p)$, Eq.~\eqref{eqS:sKLDbound} can be written as
\begin{align}
	h_{\rm S}(d_T (\bm p, \bm q))  \leq D (\bm p,\bm q).  \label{eqS:sKLDbound_final}
\end{align}
where $h_{\rm S} (x) \equiv x \ln \frac{1+x}{1-x}$. For Eq.~\eqref{eqS:sKLDbound_final},  $\tanh^{-1} (x) = \frac{1}{2} \ln \frac{1+x}{1-x}$ is used.

\begin{figure}
	\includegraphics*[width=0.6\columnwidth]{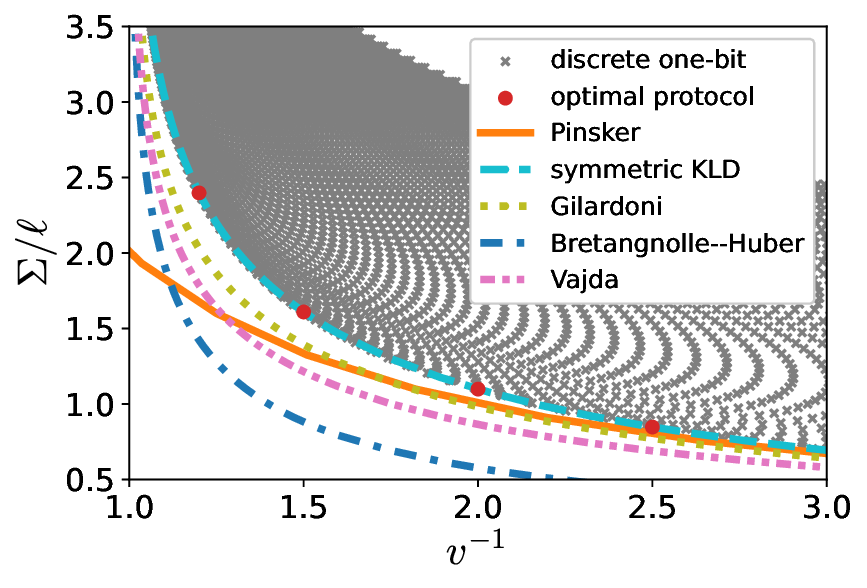}
	\caption{Comparison between various bounds. The symmetric KLD bound is the tightest: the result of the optimal protocol is on the symmetric KLD bound.} \label{figS:comparison}
\end{figure}

\section{Optimal protocol for a single-bit system}
We consider the time evolution of a probability distribution $\bm p(t) = \{ p_0(t), p_1 (t) \}$ at time $t$ for a single-bit (two-state) system
\begin{align}
	\dot p_0 (t) &= R_{01}(t) p_1 (t) - R_{10}(t) p_0(t), \nonumber \\
	\dot p_1 (t) &= R_{10}(t) p_0 (t) - R_{01}(t) p_1(t). \nonumber
\end{align}
Note that $R_{nm} = R_{nm}^*$ for a two-state system and the total EP rate and the dynamical activity are written as
\begin{equation}
	\begin{aligned}
		\dot{\Sigma}(t) & = \dot{\Sigma} _{\rm HS}(t)= (R_{01}(t) p_1(t) - R_{10}(t) p_0(t)) \ln\left[ \frac{R_{01}(t)p_1(t)}{R_{10}(t) p_0(t)}\right], \\
		A(t) & = R_{01}(t) p_1(t) + R_{10}(t) p_0(t). \label{eqS:EPandActivity}
	\end{aligned}
\end{equation}
We provide an explicit form of dynamics for a bit system that saturates the equality of the following speed limit bound
\begin{equation}
	\frac{\ell}{A_{\rm tot}} \leq f\left( \frac{\Sigma}{ A_{\rm tot}} \right)\Longleftrightarrow h\left( \frac{\ell}{A_{\rm tot}} \right)  \leq \frac{\Sigma}{ A_{\rm tot}}, \label{eqS:main_speed_limit}
\end{equation}
for the tightest bound, i.e., the symmetric KLD bound  $h(x) = h_{\rm S} (x) = x \ln \frac{1+x}{1-x}$~\cite{gilardoni2006minimum, sason2014tight}. We first show that the following condition,
\begin{equation}
	\frac{d}{dt} \left[ \frac{R_{01}(t) p_1(t)}{R_{10}(t) p_0(t)} \right] = 0 \Longleftrightarrow \frac{R_{01}(t) p_1(t)}{R_{10}(t) p_0(t)} = c ~({\rm const.}),~\forall ~0\leq t \leq \tau~,
	\label{eq:Supp_cond}
\end{equation}
with monotonic change of $p_n(t)$ in time is sufficient to saturate the symmetric KLD bound.
\begin{proof}
	Equation~\eqref{eqS:main_speed_limit} is derived by using the four inequalities in the main text: Eqs.~(3), (4), (7), and (8). Thus, we should find the equality conditions for the four inequalities to find the saturation condition of Eq.~\eqref{eqS:main_speed_limit}.
	Without loss of generality, we consider the case of $p_0(\tau) \geq p_0(0)$ and $p_1(\tau) \leq p_1(0)$.
	We first observe that the inequality, Eq.~(3) of the main text,
	\begin{equation}
		\ell = \frac{1}{2} \sum_n \left | p_n(\tau) - p_n(0) \right| \leq \frac{1}{2} \int_0^\tau \sum_n |\dot p_n(t)| dt,
	\end{equation}
	is saturated when $p_0 $ ($p_1$) is monotonically increasing (decreasing) in time, i.e.,
	\begin{equation}
		\begin{aligned}
			\dot p_0(t) &= -R_{10}(t)p_0(t) + R_{01}(t) p_1(t) \geq 0, \\
			\dot p_1(t) &= R_{10}(t)p_0(t) - R_{01}(t) p_1(t) \leq 0.
		\end{aligned}
		\label{eq:Supp_monotone}
	\end{equation}
	For a bit system, we then have the following formula,
	\begin{equation}
		\frac{1}{2} \sum_n |\dot p_n(t)| = -R_{10}(t)p_0(t) + R_{01}(t) p_1(t). \label{eqS:second_inequality}
	\end{equation}
	The next inequality we consider is Eq.~(4) of the main text, that is,
	\begin{equation}
		\frac{1}{2} \sum_n |\dot p_n(t)| \leq A(t) d_T(\boldsymbol{Q}, \boldsymbol{Q^*}),
	\end{equation}
	where
	$$
	\begin{aligned}
		Q_{mn}(t) &= \frac{ (1-\delta_{nm}) R_{nm}(t) p_m(t)}{A(t)}~~~\textrm{and}~~~
		Q^*_{mn}(t) = \frac{ (1-\delta_{nm}) R_{mn}(t) p_n(t)}{A(t)},
	\end{aligned}
	$$
	for a two-state model. 	We note that this inequality is always saturated for a bit system by directly calculating the total variational distance as $d_T(\boldsymbol{Q}, \boldsymbol{Q^*}) = \frac{1}{A(t)} \left( -R_{10}(t)p_0(t) + R_{01}(t) p_1(t)\right) $ which is equivalent to $\frac{\sum_n |\dot p_n (t)|}{2A(t)}$ from Eq.~\eqref{eqS:second_inequality}. We observe that for any bit system, the third inequality (Eq.~(7) of the main text),
	$$
	h(d_T(\boldsymbol{Q},\boldsymbol{Q^*})) \leq \frac{\dot \Sigma(t)}{A(t)},
	$$
	is also saturated for the tightest bound $h_{\rm S}(x) = x \ln \frac{1+x}{1-x}$. Using Eqs.~\eqref{eqS:second_inequality} and \eqref{eqS:EPandActivity}, this can be directly shown by
	$$
	\begin{aligned}
		h_{\rm S} (d_T(\boldsymbol{Q},\boldsymbol{Q^*})) &= h_{\rm S} \left( \frac{\sum_n|\dot p_n(t)|}{2A(t)} \right)
		=h_{\rm S} \left( \frac{-R_{10}(t)p_0(t) + R_{01}(t) p_1(t) }{A(t)} \right) \\
		&= \frac{-R_{10}(t)p_0(t) + R_{01}(t) p_1(t) }{A(t)} \ln\left[ \frac{R_{01}(t) p_1(t)}{R_{10}(t)p_0(t) } \right] = \frac{\dot \Sigma(t)}{A(t)},
	\end{aligned}
	$$
	
	Finally, in order to saturate the inequality (Eq.~(8) of the main text),
	$$
	\ell = \int_0^\tau A(t)   f\left(\frac{\dot\Sigma(t)}{A(t)} \right) dt \leq A_{\rm tot}  f\left(\frac{\Sigma_{\rm tot}}{A_{\rm tot}} \right),
	$$
	for a concave function $f$, we require the condition
	$$
	\frac{\dot\Sigma(t)}{A(t)} = \frac{(R_{01}(t) p_1(t) - R_{10}(t) p_0(t)) \ln\left[ \frac{R_{01}(t)p_1(t)}{R_{10}(t) p_0(t)}\right] }{R_{10}(t)p_0(t) + R_{01}(t)p_1(t)}
	= \frac{\left(\frac{R_{01}(t) p_1(t)}{R_{10}(t) p_0(t)} - 1\right) \ln\left[ \frac{R_{01}(t)p_1(t)}{R_{10}(t) p_0(t)}\right] }{ 1 + \frac{R_{01}(t) p_1(t)}{R_{10}(t) p_0(t)} } = {\rm const.},
	$$
	which leads to Eq.~\eqref{eq:Supp_cond}. As $ \frac{R_{01}(t) p_1(t)}{R_{10}(t) p_0(t)} = c $ with $c>1$ directly meets the condition in Eq.~\eqref{eq:Supp_monotone}, we complete the proof. For the opposite case $p_0(\tau) \leq p_0(0)$, we reach the same conclusion with $c<1$.
\end{proof}

Now we find the explicit form of the dynamics of a bit system satisfying the equality of Eq.~\eqref{eqS:main_speed_limit} with the symmetric KLD bound. Equation~\eqref{eq:Supp_cond} leads to
\begin{equation}
	\dot p_0(t) = -R_{10}(t)p_0(t) + R_{01}(t) p_1(t)  = (c-1) R_{10}(t)p_0(t). \label{eqS:saturate_equation}
\end{equation}
The solution of Eq.~\eqref{eqS:saturate_equation} is
\begin{equation}
	p_0(t) = p_0(0) e^{(c-1)\int_0^t R_{10}(t') dt'}.
\end{equation}
Consequently, we have
\begin{equation}
	R_{01}(t) = c R_{10}(t) \frac{p_0(t)}{p_1(t)}  = \frac{ c R_{10}(t) p_0(0)}{e^{-(c-1)\int_0^t R_{10}(t') dt'} - p_0(0)}.
\end{equation}
The relevant physical quantities can also be simplified as
$$
\begin{aligned}
	\dot \Sigma(t) &= [R_{01}(t) p_1(t) - R_{10}(t) p_0(t)] \ln\left[ \frac{R_{01}(t)p_1(t)}{R_{10}(t) p_0(t)}\right] = (c-1) (\ln c) R_{10}(t) p_0(t)\\
	A(t) &= R_{10}(t)p_0(t) + R_{01}(t)p_1(t) = (c+1)R_{10}(t)p_0(t).
\end{aligned}
$$

For a special case of a bit erasing process $\boldsymbol{p}(0) = ( 1/2, 1/2)$ and $\boldsymbol{p}(\tau) = (1-\epsilon, \epsilon)$ with a constant transition rate  $R_{10}(t) = 1$, we have
$$
\begin{aligned}
	p_0(t) &= \frac{1}{2} e^{(c-1) t}, \\
	R_{01}(t) &= \frac{ 2c }{2e^{-(c-1) t} - 1},
\end{aligned}
$$
and the entropy production rate and the dynamical activity become

\begin{align}
	\dot \Sigma(t) &= (c-1) (\ln c) p_0(t) = \frac{(c-1) (\ln c)}{2} e^{(c-1) t},\\
	A(t) &= (c+1)p_0(t) = \frac{c+1}{2} e^{(c-1) t}.
	\label{eqS:Atot}
\end{align}

In order to reset the memory with error $\epsilon$, the required process time $\tau$ is given as
\begin{equation}
	p_0(\tau) = 1 - \epsilon = \frac{1}{2}e^{(c-1)\tau} \Longrightarrow \tau = \frac{\ln \left[2(1-\epsilon)\right]}{c-1}. \label{eqS:tau}
\end{equation}
Then, the total entropy production throughout the process is given as
\begin{align}
	{\Sigma} &= \int_0^{\tau} \dot{\Sigma}(t) dt \nonumber \\
	&= \frac{(c-1) (\ln c)}{2} \int_0^{\tau} e^{(c-1)t} dt \nonumber \\
	&= \left( \frac{\ln{c}}{2} \right) (1- 2\epsilon), \label{eqS:totEP}
\end{align}
or equivalently, ${\Sigma} = \frac{ (1- 2\epsilon) \ln{ \left[\frac{\ln(2(1-\epsilon))}{\tau}+1\right]}}{2}$. From Eqs.~\eqref{eqS:Atot} and \eqref{eqS:tau}, the total activity becomes
\begin{align}
	A_{\rm tot} = \int_0^\tau A(t) dt = \frac{c+1}{2(c-1)}(1-2\epsilon),
\end{align}
which leads to $v^{-1} = A_{\rm tot} / \ell = (c+1)/(c-1)$ with $c>1$. For fixed $\epsilon$, $v^{-1} \rightarrow 1^+$ is achieved in $c \rightarrow \infty$ limit, resulting in $\tau \rightarrow 0$ from Eq.~\eqref{eqS:tau}. This indicates that the process should be highly irreversible and fast for approaching $v^{-1} \rightarrow 1^+$ limit. On the other hand, $v^{-1} \rightarrow \infty$ is attained via a nearly reversible process $(c \rightarrow 1^+)$ with very large $\tau$, which is the quasistatic process. The result of the optimal protocol is plotted in Fig.~\ref{figS:comparison} and Figs.~1(c) and 1(d) of the main text.

\section{Landauer's bound for a coarse-grained bit system}

Here we show the entropy production (EP) of an original system is always larger than or equal to the EP of the associated coarse-grained bit system and  also satisfies the finite-time Landauer's bound derived in the main text. We follow the derivation in Ref.~\cite{Zhen}. Consider a physical system consisting of $N$ discrete states, where the probability of each state $n$ is denoted by $p_n (t)$ for $n \in \{1, 2, \cdots , N\}$. Dynamics of the system is described by the following master equation:
\begin{align}
	\dot p_n (t) = \sum_{m (\neq n)} \left[ R_{nm} (t) p_m(t) - R_{mn} p_n(t) \right],
\end{align}
where $R_{nm}(t)$ is the transition rate from state $m$ to $n$. The EP rate $\dot \Sigma(t)$ of this system is written as
\begin{align}
	\dot \Sigma(t) = \frac{k_{\rm B}}{2} \sum_{n \neq m} \left[ R_{nm} (t) p_m(t) - R_{mn} (t) p_n(t) \right] \ln   \left[ \frac{R_{nm} (t) p_m(t)}{R_{mn} (t) p_n(t)} \right], \label{eqS:EPoriginal}
\end{align}
where $k_{\rm B}$ is the Boltzmann constant.

These $N$ microstates can be divided into two  sets, $\Omega_0$ and $\Omega_1$. Then the system can be treated as a coarse-grained bit system by regarding the system being in state ``0'' (``1'') when its microstate belongs to $\Omega_0$ ($\Omega_1$). The probability of this coarse-grained bit system $p_a^{\rm cg} (t)$ for $a\in \{0, 1\}$ is given by
\begin{align}
	p_a^{\rm cg} (t) = \sum_{i \in \Omega_a} p_i (t).
\end{align}
The master equation for the coarse-grained probability $p_a^{\rm cg}(t)$ can be derived as follows:
\begin{align}
	\dot p_a^{\rm cg} =\sum_{n \in \Omega_a} \dot p_i (t) &= \sum_{n \in \Omega_a} \sum_{m(\neq n)} \left[R_{nm}(t)p_m(t) - R_{mn} (t) p_n(t) \right] \nonumber \\
	&= \sum_{n \in \Omega_a} \left(  \sum_{m \in \Omega_{\bar a}} +\sum_{m(\neq n) \in \Omega_a} \right) \left[R_{nm}(t)p_m(t) - R_{mn} (t) p_n(t) \right] \nonumber \\
	&= \sum_{n \in \Omega_a} \sum_{m \in \Omega_{\bar a}} \left[R_{nm}(t) \frac{p_m(t)}{p_{\bar a}^{\rm cg} (t)} p_{\bar a}^{\rm cg} (t) - R_{mn} (t) \frac{ p_n(t)}{p_a^{\rm cg} (t)} p_a^{\rm cg} (t) \right] \nonumber \\
	& = R_{a \bar a}^{\rm cg}(t) p_{\bar a}^{\rm cg}(t) - R_{\bar a a}^{\rm cg}(t) p_a^{\rm cg}(t), \label{eqS:bit_mater_eq}
\end{align}
where $R_{a \bar a}^{\rm cg} (t) = \sum_{n \in \Omega_a} \sum_{m \in \Omega_{\bar a}} R_{nm} (t) p_m (t)/ p_{\bar a}^{\rm cg}(t) $. For deriving the third equality of Eq.~\eqref{eqS:bit_mater_eq}, we use
\begin{align}
\sum_{n \in \Omega_a} \sum_{m(\neq n) \in \Omega_a} \left[R_{nm}(t)p_m(t) - R_{mn} (t) p_n(t) \right] = \sum_{n \neq m \in \Omega_a}R_{nm}(t)p_m(t) -  \sum_{n \neq m \in \Omega_a}  R_{mn} (t) p_n(t) = 0. \label{eqS:intra_transition}
\end{align}
The first summation in Eq.~\eqref{eqS:intra_transition} is over intra-transitions taking place inside the same $\Omega_a$, which has no contribution to the change of $p_a^{\rm cg}$, thus it should be zero.   From the master equation of $p_a^{\rm cg}$, Eq.~\eqref{eqS:bit_mater_eq}, we can write the expression of the EP rate $\dot \Sigma^{\rm cg}(t)$ for the coarse-grained bit system as
\begin{align}
	\dot \Sigma^{\rm cg} (t) = \frac{k_{\rm B}}{2} \sum_{a} \left[ R_{a \bar a}^{\rm cg} (t) p_{\bar a}^{\rm cg} (t) - R_{\bar a a}^{\rm cg}(t) p_a^{\rm cg}(t) \right] \ln   \left[ \frac{R_{a \bar a}^{\rm cg} (t) p_{\bar a}^{\rm cg}(t)}{R_{\bar a a}^{\rm cg} (t) p_a^{\rm cg} (t)} \right].
\end{align}

The EP of the original system~\eqref{eqS:EPoriginal} can be separated into two parts as follows:
\begin{align}
	\dot \Sigma(t) &= \frac{k_{\rm B}}{2} \sum_{a} \left( \sum_{n \in \Omega_a} \sum_{m \in \Omega_{\bar a}} + \sum_{n \neq m \in \Omega_a} \right) \left[ R_{nm} (t) p_m(t) - R_{mn} (t) p_n(t) \right] \ln   \left[ \frac{R_{nm} (t) p_m(t)}{R_{mn} (t) p_n(t)} \right] \nonumber \\
	& = \dot \Sigma^{\rm inter}(t) + \dot \Sigma^{\rm intra} (t),
\end{align}
where $\dot \Sigma^{\rm inter}(t)$ is the EP induced by inter-transitions between microstates $n$ and $m$ belonging to different sets such as $n \in \Omega_a $ and $m \in \Omega_{\bar a} $ for $a \neq \bar a$ defined as
\begin{align}
	\dot \Sigma^{\rm inter}(t) \equiv \frac{k_{\rm B}}{2} \sum_{a}  \sum_{n \in \Omega_a} \sum_{m \in \Omega_{\bar a}}  \left[ R_{nm} (t) p_m(t) - R_{mn} (t) p_n(t) \right] \ln   \left[ \frac{R_{nm} (t) p_m(t)}{R_{mn} (t) p_n(t)} \right]
\end{align}
and $\Sigma^{\rm intra} (t)$ is the EP induced by intra-transitions between states belonging to the same $\Omega_a$ defined as
\begin{align}
	\dot \Sigma^{\rm intra}(t) \equiv \frac{k_{\rm B}}{2} \sum_{a}  \sum_{n \neq m \in \Omega_a}  \left[ R_{nm} (t) p_m(t) - R_{mn} (t) p_n(t) \right] \ln   \left[ \frac{R_{nm} (t) p_m(t)}{R_{mn} (t) p_n(t)} \right].
\end{align}
We can easily check that $\dot \Sigma^{\rm inter} (t)$ and $\dot \Sigma^{\rm intra} (t)$ are non-negative from the inequality $(a-b)\ln (a/b) \geq 0$ for $a, b>0$. Therefore,  we  get
\begin{align}
	\dot \Sigma (t) &= \dot \Sigma^{\rm inter} (t) + \dot \Sigma^{\rm intra} (t) \geq \dot \Sigma^{\rm inter} (t) \nonumber \\
	& \geq  \frac{k_{\rm B}}{2} \sum_{a} \left[ \sum_{n \in \Omega_a} \sum_{m \in \Omega_{\bar a}}  \left\{ R_{nm} (t) p_m(t) - R_{mn} (t) p_n(t) \right\} \ln   \left\{\frac{\sum_{n \in \Omega_a} \sum_{m \in \Omega_{\bar a}}R_{nm} (t) p_m(t)}{\sum_{n \in \Omega_a} \sum_{m \in \Omega_{\bar a}}R_{mn} (t) p_n(t)} \right\}  \right] \nonumber \\
	& = \frac{k_{\rm B}}{2} \sum_a	\left( R_{a \bar a}^{\rm cg} (t) p_{\bar a}^{\rm cg}(t) - R_{\bar a a}^{\rm cg} (t) p_{a}^{\rm cg}(t) \right) \ln \frac{R_{a\bar a}^{\rm cg}(t) p_{\bar a}^{\rm cg}(t) }{R_{\bar a a}^{\rm cg}(t) p_{a}^{\rm cg}(t)} \nonumber \\
	&= \dot \Sigma^{\rm cg} (t). \label{eqS:EPrate_inequality}
\end{align}
The log-sum inequality, $\sum_n u_n \ln (u_n/v_n) \geq (\sum_n u_n) \ln (\sum_n u_n/ \sum_n v_n)$ for any $u_n, v_n \geq 0$, is used to derive the second line of Eq.~\eqref{eqS:EPrate_inequality}.
From Eq.~\eqref{eqS:EPrate_inequality}, we have $\Sigma(\tau) - \Sigma(0) \geq \Sigma^{\rm cg}(\tau) - \Sigma^{\rm cg}(0)$. As we can set $\Sigma(0)=\Sigma^{\rm cg}(0)=0$, we finally have the inequality:
\begin{align}
	\Sigma(\tau) \geq \Sigma^{\rm cg} (\tau).
\end{align}
As $\Sigma(\tau)$ is always larger than or equal to $\Sigma^{\rm cg} (\tau)$, $\Sigma(\tau)$ is also bounded by the finite-time Landauer's bound derived in the main text as
\begin{align}
	\Sigma(\tau) \geq \Sigma^{\rm cg} (\tau) \geq \ell^{\rm cg} \frac{h(v^{\rm cg})}{v^{\rm cg}} \equiv B_h(\ell^{\rm cg}, v^{\rm cg}),
\end{align}
where $\ell^{\rm cg}$ and $v^{\rm cg}$ are distribution distance  and jump distance, respectively, evaluated in the coarse-grained bit system. It is worthwhile to note that $\Sigma(\tau)$ is not tightly bounded by $B_h(\ell^{\rm cg}, v^{\rm cg})$ as long as $\Sigma^{\rm intra}(\tau)$ is not negligible. The above derivation for a discrete system is straightforwardly applicable to a continuous case~\cite{Zhen}.

\bibliography{speed}

